\documentclass[12pt]{amsart}
\usepackage{amsmath}
\usepackage{amsthm}
\usepackage{amsfonts}
\usepackage{graphicx}
\usepackage{amssymb} 
\usepackage{bbold}  
\newenvironment{nouppercase}{
  
  \renewcommand{\uppercasenonmath}[1]{}}{}

\begin{document}

\title[Quantum States from Minimal Surfaces]
{Quantum States from Minimal Surfaces}
\author[Jens Hoppe]{Jens Hoppe}
\address{Braunschweig University, Germany}
\email{jens.r.hoppe@gmail.com}

\begin{abstract}
Apart from relating interesting quantum--mechanical systems
to equations describing a parabolic discrete minimal surface,
the quantization of a cubic minimal surface in $\mathbb{R}^4$ is considered.
\end{abstract}

\begin{nouppercase}
\maketitle
\end{nouppercase}
\thispagestyle{empty}
\noindent
Some time ago, on the last 5 pages following some more general considerations of Quantum Minimal Surfaces \cite{AHK19}, a non-linear recursion relation,
\begin{equation}\label{eq1} 
v_{n+1} = \dfrac{(n+1)\varepsilon}{v_n} - v_{n-1} - 1, \quad v_{-1} = 0,
\end{equation}
arising from ($(W^{\dagger} W)_{nn} = v_n > 0$)
\begin{equation}\label{eq2} 
[W^{\dagger}, W] + [W^{\dagger^2}, W^2] = \varepsilon
\end{equation}
(cp. \cite{CT99}), a quantum analogue of a parabolic $(x_3 + ix_4 = (x_1 + ix_2)^2)$ minimal surface in $\mathbb{R}^4$ \cite{K1897, E1912}, 
\begin{equation}\label{eq3} 
\vec{x}(r,u) = (r \cos u, r\sin u, r^2 \cos 2u, r^2 \sin 2u),
\end{equation}
was shown to lead to polynomial functions $u_n(x)$, resp. (via $v_n = \dfrac{u_n}{u_{n-3}}\dfrac{u_{n-4}}{u_{n-1}} = \dfrac{\tau_{n+1}\tau_{n-1}}{\tau^2_n}$) polynomials
\begin{equation}\label{eq4} 
\tau_n(x) = u_{n-1}u_{n-2}u_{n-3},
\end{equation}
for $n > 2$ of degree $(n+1)$ in $x := v_0$ $(\tau_1 = x, \tau_2 = x(\varepsilon-x))$, indicating `integrability'.\\\\
In 2021, I noticed the consistency of \eqref{eq1} with the existence of polynomials $P_k(z)$ (of degree $k$, parity and sign $(-)^k$; $P_0 = 1, P_1 = -2z, P_2 = 4(2z^2 + 1), P_3 = -8(5z^3 + 9z)$),
\begin{equation}\label{eq5} 
v_n = (n+1)\varepsilon \sum_{k=0}^{\infty}\varepsilon^k P_k(n+1)
\end{equation}
\begin{equation}\label{eq6} 
-P_{k+1}(z) = \sum_{i+j=k} P_j(z)[(z+1)P_i(z+1) + (z-1)P_i(z-1)]
\end{equation}
(and in 2023 contacted Andrew Hone, who immediately had some insights concerning the problem). While the existence of a positive solution of \eqref{eq1} was proven
in \cite{AHK19}, its assumed uniqueness and explicit form remained open, until recently; early this year A. Hone told me that he and his collaborators to a large extent explicitly solved \eqref{eq1} by relating it to Painlev\'e V, which for special values of the parameters has elementary solutions related to Bessel functions and Riccati equations.\\
Going back to thinking about the problem I found some elementary consequences of \eqref{eq5} (including concerning their positivity), interesting Schr\"odinger potentials related to \eqref{eq1}, and a simple proof that if $v_0 = v_0(s := \frac{1}{6\varepsilon})$ satisfies $(n=0)$ 
\begin{equation}\label{eq7} 
\dfrac{1}{2}v'_n = k_n v_n^2 - f_n(s)v_n - g_n(s) 
\end{equation}
(with\footnote{then uniquely transporting $k_n = 1$} $k_0 = 1$ $f_0 = -1 + \frac{2}{6s}$ $g_0 = \frac{1}{6s}$) all higher $v_n$ will (as a consequence of \eqref{eq1}) -- with $k_n = 1$,
\begin{equation}\label{eq8} 
f_n = -1 + \varepsilon \gamma_0^{(n)} - 2 \sum_{i=2}^n (-)^i v_{n-i}, \quad g_n = \dfrac{n+1}{6s},
\end{equation}
$\gamma_0^{(n)}=2$ for even $n$ and $\gamma_0^{(n)} = 1$ for odd $n$.\\\\
\eqref{eq7}/\eqref{eq8} can be proven by induction as follows: 
\begin{equation}\label{eq9} 
\begin{split}
\frac{1}{2}v'_{n+1} & = -\frac{1}{2s}\big[ \frac{(n+1)}{6sv_n} = v_{n+1} + v_{n-1} + 1\big] \\[0.15cm]
& \hphantom{==} + \big[ -\frac{n+1}{6sv_n^2} (\frac{1}{2}v'_n = k_n v_n^2 - f_n v_n - g_n) \\[0.15cm]
& \hphantom{= + \big[ - =} = -\frac{n+1}{6s}k_n + f_n (v_{n+1} + v_{n-1}+1) \\[0.15cm]
& \hphantom{= + \big[ - ======} + \frac{6s}{n+1}(v_{n+1} + v_{n-1} +1)^2 g_n \big]\\[0.15cm]
& = \big( \frac{6s}{n+1}g_n \big)v^2_{n+1} - \big[f_{n+1} = \frac{1}{2s}- f_n - 2\frac{6s}{n+1}g_n(v_{n-1}+1)\big]v_{n+1} - g_{n+1} \\[0.35cm]
- g_{n+1} & = v_{n-1}\big[2(k_{n+1}-k_n) + (k_{n+1}-k_{n-1})v_{n-1} - 2k_nv_{n-2}\big] \\[0.15cm]
& \hphantom{==} -\frac{1}{2s} - \frac{k_n}{6s} + f_n + k_{n+1} = \ldots = -\frac{(n+2)}{6s}
\end{split}
\end{equation}
(using $k_n = 1, g_n = \frac{n+1}{6s}$).\\\\
Solutions of \eqref{eq7} (for general $g_n >0$; from now on we will take/assume the concrete/actual case : $k_n = 1, g_n = \frac{n+1}{6s}$) have the property that, if positive for large $s$ (like those of the form \eqref{eq5}) have to remain positive (for all $s > 0$), as if $v_n = 0$, $v'_n = -g_n <0$ at that point -- which is a contradiction, as if coming from the right, zero can only be reached with $v'_n \geqslant  0$. Also, 
\begin{equation}\label{eq10} 
f_n = -1 + \varepsilon\sum_0^{\infty} \varepsilon^l \gamma_l (z = n+1)
\end{equation}
may actually be calculated, for all $n$, recursively, as inserting \eqref{eq10} and \eqref{eq5} into \eqref{eq7} $(k_n = 1,\, g_n = \frac{n+1}{6s})$ gives
\begin{equation}\label{eq11} 
-P_{k+1}(z) = 3(k+1)P_k + z\sum_{i+j=k} P_i(z)P_j(z) - \sum_{i+l=k} \gamma_l (z) P_i(z)
\end{equation}
i.e.
\begin{equation}\label{eq12} 
\begin{split}
\gamma_0(z) & = -(z-3),\qquad \gamma_1(z) = 2(z-1)(z-2), \\[0.15cm]
\gamma_2(z) & = -4(z-1)(z-2)(2z-3), \\[0.15cm]
\gamma_3(z) & = 8(z-1)(z-2)(5z^2 - 15z+14).
\end{split}
\end{equation}
Finally note that \eqref{eq10}, together with the recursion in \eqref{eq9} for $f_n$, implies 
\begin{equation}\label{eq13} 
\begin{array}{l}
\gamma_{k>0}(z+1) + \gamma_k(z) = -2(z-1) P_k(z-1)\\[0.25cm]
\gamma_0(z+1) + \gamma_0(z) = 3-2(z-1)
\end{array}
\end{equation}
i.e. in particular $\gamma_{k>0}(1) + \gamma_{k>0}(2) = 0$, and -- iterating \eqref{eq13} --
\begin{equation}\label{eq14} 
 -\frac{1}{2}\gamma_{k>0}(z=n+1) = (n-1)P_k(n-1)-(n-2)P_k(n-2) \pm \ldots
\end{equation}
Writing $v_n = -\frac{1}{2} \frac{\phi'_n}{\phi_n}$ in \eqref{eq7} on the other hand gives
\begin{equation}\label{eq15} 
\phi''_n + 2f_n\phi'_n - 4\frac{n+1}{6s} \phi_n = 0;
\end{equation}
so, with
\begin{equation}\label{eq16} 
 \psi_n = G_n \phi_n, \quad G'_n = f_nG_n
\end{equation}
\begin{equation}\label{eq17} 
 -\psi''_n + W_n \psi_n = -\psi_n
\end{equation}
where the potential in the one--dimensional Schr\"odinger equation is given by
\begin{equation}\label{eq18} 
\begin{split}
W_n & = f'_n + f^2_n - 1 + 4\frac{n+1}{6s}\\[0.15cm]
W_0 & = -\frac{2}{9s^2}, \quad W_1 = -\frac{5}{36s^2} +\frac{1}{s}, \quad W_2 = - \frac{2}{9s^2} + \frac{2}{s},
\end{split}
\end{equation}
corresponding to
\begin{equation}\label{eq19} 
\begin{split}
f_0 & = - 1 + \frac{2}{6s}, \quad f_1 = -1 + \frac{1}{6s}, \\[0.15cm]
f_2 & = - 1 + \frac{2}{6s} -2v_0, \quad f_3 = -1 + \frac{1}{6s} -2v_1 + 2v_0 \\[0.15cm]
G_0 & = e^{-s}s^{\frac{1}{3}}, \quad G_1 = e^{-s}s^{\frac{1}{6}}, \quad G_2 = \psi_0 = \sqrt{s}K_{\frac{1}{6}}(s). 
\end{split}
\end{equation}
Starting with $W_3$, the potential will no longer be a linear combination of $\frac{1}{s^2}$ and $\frac{1}{s}$, but involve the $v'_ns$, which according to \eqref{eq5} go like $\frac{n+1}{6s}$ at infinity, but (in the asymptotic expansion) involving also arbitrarily high power of $\frac{1}{s}$ (e.g. $W_3 = -\frac{5}{36s^2} + \frac{5}{3s} + 8v_0^2 + 8v_0 - \frac{2}{3s}v_0$).\\ 
The potential $W_0 = -\frac{2}{9s^2}$ is particularly interesting, as it is tempting to take $\psi_0 = \sqrt{s}K_{\frac{1}{6}}(s)$ (which is finite, positive and vanishes both at $s=0$ and at $\infty$) as the ground state of
\begin{equation}\label{eq20} 
H_0 = -\partial^2 - \frac{2}{9s^2}.
\end{equation}
Taken without any further comments/modifications, this can not be true, as \eqref{eq20} is scale invariant and by taking $\psi = \psi_0(\lambda s)$ any negative `Eigenvalue' could be produced (from which it follows that \eqref{eq20} can not be self-adjoint\footnote{many thanks to J. Fr\"ohlich for several discussions on this, in particular pointing out that $\psi_0(s)$ and $\psi_0(\lambda s)$ (both being positive on $(0,+\infty)$) can not be orthogonal to each other}). Some comments about the subtleties can be found e.g. in \cite{EG2006}; note however 
\begin{equation}\label{eq21} 
\begin{split}
\int_0^{\infty} \psi (-\partial^2 - \frac{2}{9s^2})\psi & = \int_0^{\infty} \psi (\partial + \frac{1}{3s})(-\partial + \frac{1}{3s}) \psi \\[0.15cm]
& = \int_0^{\infty} \psi(\partial+\frac{1}{3s})(\frac{\psi}{3s} - \psi')\\[0.15cm]
& = \int_0^{\infty} \big(-\psi' + \frac{\psi}{3s}\big)^2 ds + \big[ \psi(\frac{\psi}{3s} - \psi') \big]_0^{\infty}. 
\end{split}
\end{equation}
Although the leading singularity of $\psi'(s\rightarrow 0)$ is precisely cancelled by $\frac{\psi}{3s}$, there is a subleading singularity\footnote{many thanks to T. Turgut for going with me through the argument and spotting the (effect of the) subleading term in $K_{\frac{1}{6}}$} in the Bessel function derivative which makes the boundary term in \eqref{eq21} finite (negative), rather than it being zero.\\
Concerning the observed polynomiality of the $u_n$ and $\tau_n$ (cp.[1]), consider 
\begin{equation}\label{eq22} 
\begin{split}
\tau_{n+1} & = r_n\tau_n = r_n r_{n-1}\cdot \ldots \cdot r_0 = u_n u_{n-1}u_{n-2} \\[0.15cm]
& = v_0v_1\ldots v_n \tau_n = v_0^{n+1}v_1^n \ldots v_{n-1}^2v_n
\end{split}
\end{equation}
and the following consequence of \eqref{eq1}: 
\begin{equation}\label{eq23} 
\begin{split}
\tau_{n+2} & = \tau_{n+1}v_0v_1 \ldots v_{n-1}\big[ v_n v_{n+1}  = \varepsilon_n - v_n(v_{n-1} + 1)\\[0.15cm]
& \hphantom{===============} = \varepsilon_n - v_n\big( \frac{\varepsilon_{n-2}}{v_{n-2}} - v_{n-3}\big) \big]\\[0.15cm]
& = \tau_{n+1} \big[ \varepsilon_nr_{n-1} - r_n\big( \varepsilon_{n-2}\frac{r_{n-3}}{r_{n-2}} - \frac{r_{n-3}}{r_{n-4}} \big) \big]\\[0.15cm]
& = u_n u_{n-1} u_{n-2}\big[ \varepsilon_n\frac{u_{n-1}}{u_{n-4}} -\varepsilon_{n-2}\frac{u_n}{u_{n-6}}\frac{u_{n-5}}{u_{n-2}} + \frac{u_n}{u_{n-6}}\frac{u_{n-7}}{u_{n-4}}\big] \\[0.15cm]
& = \frac{u_{n-1} u_n}{u_{n-4}u_{n-6}}\big[ \varepsilon_n u_{n-2} u_{n-1} u_{n-6} - \varepsilon_{n-2} u_n u_{n-5} u_{n-4} + u_nu_{n-7} u_{n-2}\big] \\[0.15cm]
& \stackrel{!}{=} u_{n-1}u_n u_{n+1};
\end{split}
\end{equation}
so, (assuming that the $n$'s have no common factor)
\begin{equation}\label{eq24} 
\begin{split}
\varepsilon_n u_{n-2}u_{n-1}u_{n-6} + u_nu_{n-7}u_{n-2} & =: u_{n-2} u_{n-4}p_{n-3}\\[0.15cm]
u_n u_{n-7} u_{n-2} - \varepsilon_{n-2}u_n u_{n-5}u_{n-4} & =: u_{n-6}u_nQ_{n-3} 
\end{split}
\end{equation}
and, inserting \eqref{eq24} into the last part of \eqref{eq23},
\begin{equation}\label{eq25} 
\begin{split}
u_n(Q_{n-3} =: -q_{n-3}) + \varepsilon_n u_{n-2} u_{n-1} & = u_{n-4} u_{n+1} \\[0.15cm]
u_{n-2}p_{n-3}- \varepsilon_{n-2} u_n u_{n-5} & = u_{n-6}u_{n+1}
\end{split}
\end{equation}
$-$ from which it also follows that $(\varepsilon_n := \varepsilon(n+1))$
\begin{equation}\label{eq26} 
\begin{split}
u_{n+1}u_{n-4}u_{n-6} & + \varepsilon_{n-2}u_n u_{n-5}u_{n-4}\\
& = \varepsilon_n u_{n-1}u_{n-2}u_{n-6}+u_n u_{n-7} u_{n-2}, 
\end{split}
\end{equation}
in addition to the `original' (cp.[1]) identity following from \eqref{eq1},
\begin{equation}\label{eq27} 
\begin{split}
\varepsilon_{n+2}u_{n+1} u_n u_{n-1} & = u_{n+3} u_{n-1}u_{n-2}\\
& \quad + u_{n+2} \big[u_{n+1}u_{n-3}+u_n u_{n-2}\big]. 
\end{split}
\end{equation}
Writing the  three $q$--equations, and  two $p$--equations, in a slightly more convenient form: 
\begin{equation}\label{eq28} 
\begin{split}
q_n u_n & = u_{n+2} u_{n-2} + u_{n+1} u_{n-1}\\[0.15cm]
q_n u_{n+3} & = \varepsilon_{n+3} u_{n+2} u_{n+1} - u_{n+4} u_{n-1}\\[0.15cm]
q_n u_{n-3} & = \varepsilon_{n+1} u_{n-1} u_{n-2} - u_{n+1} u_{n-4}
\end{split}
\end{equation}
(the first and the last following from the 2 possible factorisations of \eqref{eq26}, and the middle one appearing in \eqref{eq24}),
\begin{equation}\label{eq29} 
\begin{split}
p_n u_{n-1} & = \varepsilon_{n+3} u_{n+2} u_{n-3} + u_{n+3} u_{n-4}\\[0.15cm]
p_n u_{n+1} & = \varepsilon_{n+1} u_{n+3} u_{n-2} + u_{n-3} u_{n+4};
\end{split}
\end{equation}
hence also
\begin{equation}\label{eq30} 
\begin{split}
\varepsilon u_n u_{n-1} & = q_{n+1} u_{n-2} - u_{n+1}q_{n-2}\\[0.15cm]
3\varepsilon u_{n+2} u_{n-3} & = p_n u_{n-1} - p_{n-1} u_n;
\end{split}
\end{equation}
finally 
\begin{equation}\label{eq31} 
\begin{split}
q_n & = \frac{1}{u^2_{n+3}}\big[u_n (u_{n+5}u_{n+1} + u_{n+4}u_{n+2}) -u_{n+3}\varepsilon u_{n+1} u_{n+2} \big] \\[0.15cm]
& =  \frac{1}{u^2_{n-3}}\big[u_n (u_{n-1}u_{n-5} + u_{n-2}u_{n-4}) + \varepsilon u_{n-1} u_{n-2}u_{n-3} \big].
\end{split}
\end{equation}
From the above one can deduce that
\begin{equation}\label{eq32} 
\begin{pmatrix}
q_{n+1} \\[0.25cm] u_{n+2}
\end{pmatrix} 
= 
\begin{pmatrix}
\frac{u_{n+1}}{u_{n-2}} & \varepsilon \frac{u_n}{u_{n-2}}\\[0.25cm]
-\frac{u_{n+1}}{u_{n-3}} & \varepsilon \frac{(n+2)u_n}{u_{n-3}}
\end{pmatrix}_{=: T_n}
\begin{pmatrix}
q_{n-2} \\[0.25cm] u_{n-2}
\end{pmatrix} 
\end{equation}
and 
\begin{equation}\label{eq33} 
\begin{pmatrix}
p_n \\[0.25cm] u_{n+3}
\end{pmatrix} 
= 
\begin{pmatrix}
\frac{u_n}{u_{n-1}} & 3\varepsilon \frac{u_{n-3}}{u_{n-1}}\\[0.25cm]
-\frac{u_n}{u_{n-4}} & -(n+1)\varepsilon \frac{u_{n-3}}{u_{n-4}}
\end{pmatrix}_{=: S_n}
\begin{pmatrix}
p_{n-1} \\[0.25cm] u_{n+2}
\end{pmatrix} 
\end{equation}
both of which can be iterated, giving products of `transfer--operators' $T_n$ (resp. $S_n$) acting on some start vector, e.g. ${q_{-1} \choose u_0} = {\varepsilon \choose x}$.\\\\
One finds
\begin{equation}\label{eq34} 
\begin{array}{l}
u_0 + u_1 = q_{-1} = \varepsilon, \quad q_0 = x+\varepsilon = \frac{u_1 + u_2}{u_0}\\[0.25cm]
q_2 = \varepsilon\big[ (1+2\varepsilon)x - \varepsilon \big], \quad q_3 = \varepsilon (-5x^2 - 2x(1+\varepsilon) + 3\varepsilon^2 + 2\varepsilon)\\[0.25cm]
p_0 = 3\varepsilon\big[ x(1+2\varepsilon) - \varepsilon \big] = 3q_2\\[0.25cm]
p_1 = -3x(6\varepsilon^2 + \varepsilon) + (18\varepsilon^2 - 5\varepsilon + 2\varepsilon^3),\: xp_1 = 5\varepsilon u_3 + u_4\\[0.25cm]
p_2(\varepsilon - x) = p_2u_1 = 6\varepsilon u_4 + u_5\\[0.25cm]
\Rightarrow p_2 = 3\varepsilon\big[ (1+6\varepsilon)x^2 + (1+10\varepsilon + 6\varepsilon^2)x - \varepsilon(9\varepsilon +1) \big].
\end{array}
\end{equation}
Finally I would like to make some remarks about the quantization of the minimal surface in $\mathbb{R}^4$ given by
\begin{equation}\label{eq36} 
\vec{x}(r,u) = (r\cos u, r\sin u, r^3\cos 3u, r^3\sin 3u).
\end{equation}
\begin{equation}\label{eq37} 
[W^{\dagger}, W] + [W^{\dagger^3}, W^3] = \varepsilon \mathbb{1}
\end{equation}
(cp. 7.42 of [1], $W|n\rangle = w_n|n+1\rangle$) gives $(v_n=|w_n|^2 \geqslant 0)$
\begin{equation}\label{eq38} 
v_n - v_{n-1} + v_n v_{n+1}v_{n+2} - v_{n-1} v_{n-2}v_{n-3} = \varepsilon, \quad v_{-1} =0,
\end{equation}
resp.
\begin{equation}\label{eq39} 
v_n(v_{n+1}v_{n+2}+v_{n-1}v_{n-2}+v_{n+1}v_{n-1}+1) = \varepsilon(n+1) = \varepsilon_n,
\end{equation}
\begin{equation}\label{eq40} 
v_{n+2} = \frac{1}{v_{n+1}}\big( \frac{\varepsilon_n}{v_n} -1 \big) - v_{n-1} \big( \frac{v_{n-2}}{v_{n+1}} + 1 \big), \quad v_{-1} = 0.
\end{equation}
In analogy with \eqref{eq5}/\eqref{eq6} one consistently finds
\begin{equation}\label{eq41} 
v_n = (n+1)\varepsilon \big( 1+ \varepsilon^2 P_1(n+1) + \varepsilon^4 P_2(n+1) + \ldots \big)
\end{equation}
\begin{equation}\label{eq42} 
\begin{split}
-P_{l+1}(z) & := \sum_{i+j+k=l} P_k(z)\big[ P_i(z+1)P_j(z-1)(z^2-1)\\
& \qquad \qquad\qquad\qquad + P_i(z+1)P_j(z+2)(z+1)(z+2)\\[0.25cm]
& \qquad \qquad\qquad\qquad + P_i(z-1)P_j(z-2)(z-1)(z-2)\big]
\end{split}
\end{equation}
$P_0(z) \equiv 1, \: P_1(z) = -3(z^2+1), \: P_2(z) = g(3z^4+22z^2 + 9)$, e.g.\\
\begin{equation}\label{eq43} 
\begin{split}
v_0 & = \varepsilon - 6\varepsilon^3 + 306\varepsilon^5 + o(\varepsilon^7)\\[0.15cm]
v_1 & = 2\varepsilon(1-15\varepsilon^2 + 9\cdot 145 \varepsilon^4 + \ldots)\\[0.15cm]
v_2 & = 3\varepsilon (1-30\varepsilon^2 + \ldots) \\[0.15cm]
v_3 & = 4\varepsilon(1-51\varepsilon^2+\ldots)
\end{split}
\end{equation}
satisfying \eqref{eq38}/\eqref{eq39}, e.g.
\begin{equation}\label{eq44} 
\begin{array}{l}
v_0(1+v_1v_2)  = \varepsilon, \\[0.25cm]
v_1(1+v_2v_3) - v_0 = \varepsilon\\[0.25cm]
v_1(1+v_2v_3+v_2v_0)  = 2\varepsilon,
\end{array}
\end{equation}
while the method outlined in \eqref{eq1} provides a semiclassical expression for the $v_n$ as follows: 
\begin{equation}\label{eq45} 
(g_{ab}) = 
\begin{pmatrix}
1+9r^4 & 0 \\
0 & r^2(1+9r^4)
\end{pmatrix}, \frac{d\tilde{r}}{dr} = \sqrt{g(r)} = r(1+9r^4),
\end{equation}
hence
\begin{equation}\label{eq46} 
\tilde{r} (r) = \frac{1}{2}r^2 + \frac{3}{2} r^6 =: \frac{1}{2\sqrt{3}}(s+s^3)
\end{equation}
up to an integration constant, chosen when replacing $\tilde{r}$ by $(n+1)\hbar$, hence (solving the cubic equation for $s$, which has one real, positive, solution)
\begin{equation}\label{eq47} 
s = \sqrt[3]{s_+}  + \sqrt[3]{s_-},\quad s_{\pm} = \sqrt{3}\tilde{r}(1\pm \sqrt{1+\frac{1}{81\tilde{r}^2}})
\end{equation}
resp., with $x:= 9\tilde{r} = 9(n+1)\hbar$
\begin{equation}\label{eq48} 
3r^2 = (x+\sqrt{1+x^2})^{\frac{1}{3}} - (\sqrt{1+x^2}-x)^{\frac{1}{3}},
\end{equation}
whose Taylor--expansion (obviously containing only even powers of $x$, consistent with \eqref{eq41}), gives twice the odd part of
\begin{equation}\label{eq49} 
\begin{split}
1+\frac{1}{3}& \big( \varepsilon + \frac{\varepsilon^2}{2} - \frac{\varepsilon^4}{8} +\frac{1}{16}\varepsilon^6 + \ldots \big) - \frac{\varepsilon^2}{9} \big(1+\frac{\varepsilon}{2} - \frac{\varepsilon^3}{8}\big)^2 \\[0.15cm]
& +\frac{5}{81}\varepsilon^3\big(1+\frac{\varepsilon}{2} - \frac{\varepsilon^3}{8} + \ldots \big)^3 - \frac{10}{3\cdot 81}\varepsilon^4\big(1+\frac{\varepsilon}{2} - \frac{\varepsilon^3}{8} + \ldots \big)^4 + \ldots,
\end{split}
\end{equation}
i.e. (semi--classically; to be compared with \eqref{eq41}/\eqref{eq43})
\begin{equation}\label{eq50} 
v_n = r^2_n = \frac{2x}{9} - 8\frac{x^3}{3^5} + \ldots = \varepsilon_n - 3\varepsilon^3_n - \frac{3}{8}27 \varepsilon^5_n +\ldots
\end{equation}
where $\varepsilon_n = (n+1)\varepsilon$ (and $\varepsilon = 2\hbar$); comparing with (\eqref{eq41}/\eqref{eq42}/\eqref{eq43}): not too bad (although the $\varepsilon_n^5$ sign--mismatch is somewhat surprising).\\\\
\textbf{Addendum}\\
One should note that the `Hamiltonian' \eqref{eq20}, leaving aside the issue of self-adjoint extension resp. problems related to the singularities, shares with the free Hamiltonian $-\partial^2$ (the $H_0$ of the standard, reflectionless, case, with eigenfunctions $e^{\pm \kappa s}$), the property that explicit solutions of the differential equation $H_0 \psi_{\kappa} = -\kappa^2 \psi_{\kappa}$ are known for {\it any} $\kappa$, namely (and similarly for other coupling constants)
\begin{equation}\label{eq35} 
\big(-\partial^2 - \frac{2}{9s^2}\big)\big( \psi_{\kappa} := \sqrt{s}B_{\frac{1}{6}}(\kappa s) \big) = -\kappa^2\psi_{\kappa},
\end{equation}
which means that the $A_{n-1}A^{\dagger}_{n-1} - \kappa^2_{n-1} = A^{\dagger}_n A_n -\kappa^2_n$ `machinery', outlined in \cite{JH92} in the context of the spectral transform and solitons of the $KdV$ equation (very popular also in Supersymmetric Quantum Mechanics; in essence going back almost 150 years, to Darboux \cite{D1882}, see e.g. \cite{Crum1955, MS1991}), can in principle be applied as follows (raising of course the question, whether and how all this may be concretely related to \eqref{eq1}): 
while figuring out the asymptotic signature of $\chi_N$ works very similar to the free case (cp.\cite{JH92}) -- meaning: for $s \rightarrow + \infty$ $\sqrt{s}B_{\frac{1}{6}}(\kappa_N s)$ will be proportional to $e^{\pm \kappa_N s}$, depending on whether one chooses $K$ or $I$ for $B$; refer to the 2 choices as (leading to) $\chi^+_N$ resp, $\chi^-_N$; so e.g. $h_1(= -\frac{\chi'_1}{\chi_1}) \sim \mp \kappa_1$, hence $\chi^{\pm}_2 = (\partial + h_1)(\sqrt{s} B^{\pm}_{(\kappa_2 s)}) \sim (\pm \kappa_2 \mp \kappa_1)^{e^{\pm\kappa_2 s}}$; choosing $\kappa_N > \kappa_{N-1} > \ldots > \kappa_1 = 1$ one thus has that $\chi^+_N \sim \prod_{i=1}^{N-1}(\kappa_N - \kappa_i)e^{+\kappa_N s}$ and $(-)^{N-1}\chi_N^+ \sim \prod_{i=1}^{N-1}(\kappa_N - \kappa_i)e^{-\kappa_N s}$ are strictly positive for large positive $s$. To show (if true) that they are non-vanishing for all positive $s$ is more complicated than in the free case, as the argument (cp. \cite{JH92}) would have to include a discussion of self-adjointness (resp. self-adjoint extensions), e.g. concerning \eqref{eq41**} involving an eigenvalue below the spectrum (of $H_{N-1}$).\\
Factorize $H_0$ as
\begin{equation}\label{eq36**} 
\big(-\partial^2-\frac{2}{9s^2}\big) = (\partial + h_0)(-\partial+h_0) -1 =: A_0A_0^{\dagger} -1,
\end{equation}
giving
\begin{equation}\label{eq37**} 
h'_0 + h_0^2 -1 = -\frac{2}{9s^2} =: W_0 
\end{equation}
with\footnote{Note the following subtlety: while \eqref{eq37**} is solved by $h_0 = \frac{\psi'_0}{\psi_0}$, $\psi_0$ being $\sqrt{s}$ times {\it any} modified Bessel function i.e. also $I_{\pm \frac{1}{6}}(s)$, those would give at infinity $h_0 \sim +1 - \frac{1}{9s^2}$, while the RHS of \eqref{eq38**}, as $v_0 \sim \frac{1}{6s}-\frac{1}{18s^2}$ at $\infty$, (according to \eqref{eq5}) gives $-1 + \frac{1}{9s^2}$, which is only produced by choosing $h_0$ to be the logarithmic derivative of $\sqrt{s}K_{\frac{1}{6}}(s)$}
\begin{equation}\label{eq38**} 
h_0 = \frac{\psi'_0}{\psi_0} = \frac{1}{2s} + \frac{K'_{\frac{1}{6}}}{K_{\frac{1}{6}}}(s) = -(2v_0 + 1) + \frac{1}{3s}
\end{equation}
as a solution on $(0, +\infty)$.\\
Inductively introduce $A_{N = 1,2, \ldots}= \partial + h_N$ via 
\begin{equation}\label{eq39**} 
H_{N-1} = A_{N-1}A^{\dagger}_{N-1} - \kappa^2_{N-1} =: A_N A_N^{\dagger} - \kappa^2_N,
\end{equation}
i.e.
\begin{equation}\label{eq40**} 
\begin{split}
W_{N-1} & = h^2_{N-1} + h'_{N-1} - \kappa^2_{N-1} \\[0.15cm]
& =  h^2_N - h'_N - \kappa^2_N \\[0.15cm]
& = W_N - 2h'_N
\end{split}
\end{equation}
which, writing $h_N = -\frac{\kappa'_N}{\kappa_N}$, i.e
\begin{equation}\label{eq41**} 
-\chi''_N + W_{N-1}\chi_N = -\kappa^2_N \chi_N
\end{equation}
hence
\begin{equation}\label{eq42**} 
\begin{split}
W_N & = W_{N-1} + 2h'_N = \ldots = 2\sum^N_{i=1}(\ln \chi_i)'' + W_0 \\
& = 2\big( \ln \big(\prod^N_{i=1}\chi_i\big) \big)'' + W_0
\end{split}
\end{equation}
and $H_N$ then defined as ($N = 1, 2, \ldots$)
\begin{equation}\label{eq43**} 
H_N = A_NA_N^{\dagger} - \kappa^2_N
\end{equation}
inductively gives
\begin{equation}\label{eq44**} 
\chi_N = A_{N-1}\cdot \ldots \cdot A_1\big(\sqrt{s}B_{\frac{1}{6}}(\kappa_N s)\big),
\end{equation}
$B$ being a modified Bessel function ($I$ or $K$, or linear combination thereof), in particular
\begin{equation}\label{eq45**} 
\chi_1 = \sqrt{s} B_{\frac{1}{6}}(\kappa_1 s)
\end{equation}
as
\begin{equation}\label{eq46**} 
\begin{split}
(-\partial^2 + W_{N-1})\chi_N & = (A_{N-1}A^{\dagger}_{N-1} - \chi^2_1)(A_{N-1} \cdot \ldots \cdot A_1) \sqrt{s}B_{\frac{1}{6}}\\[0.15cm]
& = A_{N-1}(A_{N-2} A_{N-2}^{\dagger} - \kappa^2_2)A_{N-2}\ldots A_1 \\[0.15cm]
& = A_{N-1}\ldots A_1(A_0 A_0^{\dagger} -1)\sqrt{s}B \\[0.15cm]
& = A_{N-1}\ldots A_1 H_0 \sqrt{s} B_{\frac{1}{6}} (\kappa_N s) \\[0.15cm]
& = -\kappa^2_N\chi_N.
\end{split}
\end{equation}
\textbf{Acknowledgement}:
I would like to thank J.Arnlind, I.Bobrova, J.Fr\"ohlich, A.Hone, M.Hynek, M.Kontsevich, and T.Turgut for discussions.

\end{document}